\documentclass[preprint,12pt]{elsarticle}

\usepackage{graphicx}
\usepackage{amssymb}
\usepackage{bm}
\usepackage{amsthm}
\usepackage{amsbsy}
\usepackage{amsmath}
\usepackage{amsfonts}
\usepackage{relsize}

\usepackage{caption}
\usepackage{subcaption}
\usepackage{tabularray}
\usepackage{booktabs}
\usepackage{multirow}
\usepackage{color}

\newcounter{bla}

 \def\thickhline{%
  \noalign{\ifnum0=`}\fi\hrule \@height \thickarrayrulewidth \futurelet
   \reserved@a\@xthickhline}
\def\@xthickhline{\ifx\reserved@a\thickhline
               \vskip\doublerulesep
               \vskip-\thickarrayrulewidth
             \fi
      \ifnum0=`{\fi}}

\journal{Computer Physics Communications}

\begin{document}

\begin{frontmatter}



\title{Berry: A code for the differentiation of Bloch wavefunctions from DFT calculations}


\author[a]{Leander Reascos}
\author[a]{Fábio Carneiro}
\author[d]{André Pereira}
\author[c]{Nuno Filipe Castro}
\author[a,b]{Ricardo Mendes Ribeiro\corref{author}}

\cortext[author] {Corresponding author.\\\textit{E-mail address:} ricardo@fisica.uminho.pt}
\address[a]{Department and Centre of Physics, University of Minho,
Campus of Gualtar, 4710-057, Braga, Portugal}
\address[b]{International Iberian Nanotechnology Laboratory (INL),
Av. Mestre Jose Veiga, 4715-330, Braga, Portugal}
\address[c]{Department of Physics and LIP, University of Minho,
Campus of Gualtar, 4710-057, Braga, Portugal}
\address[d]{High-Assurance Software Laboratory, INESC TEC, Rua Dr. Roberto Frias, 4200-465 Porto, Portugal}
\begin{abstract}

Density functional calculations of electronic structures of materials
is one of the most used techniques in theoretical solid state physics.
These calculations retrieve single electron wavefunctions and their eigenenergies.

The berry suite of programs amplifies the usefulness of DFT by ordering the eigenstates in analytic bands,
allowing the differentiation of the wavefunctions in reciprocal space.
It can then calculate Berry connections and curvatures and the second harmonic generation conductivity.

The berry software is implemented for two dimensional materials and was tested in hBN and InSe.
In the near future, more properties and functionalities are expected to be added.


\end{abstract}


\end{frontmatter}



{\bf PROGRAM SUMMARY}

\begin{small}
\noindent
{\em Program Title:} berry                                         \\
{\em CPC Library link to program files:} (to be added by Technical Editor) \\
{\em Developer's repository link:} https://github.com/ricardoribeiro-2020/berry \\
{\em Code Ocean capsule:} (to be added by Technical Editor)\\
{\em Licensing provisions:} MIT  \\
{\em Programming language:} Python3                                  \\
{\em Nature of problem:} Differentiation of Bloch wavefunctions in reciprocal space, numerically obtained from a DFT software, applied to two dimensional materials.
This enables the numeric calculation of material's properties such as Berry geometries and Second Harmonic conductivity.\\
{\em Solution method:} Extracts Kohn-Sham functions from a DFT calculation, orders them by analytic bands using graph and AI methods and calculates the gradient
of the wavefunctions along an electronic band.\\
{\em Applies only to two dimensional materials, and only imports Kohn-Sham functions from Quantum Espresso package.}\\
   \\

\end{small}

\section{Introduction}
\label{sec:introduction}

Ventura \emph{et al.}\cite{Ventura2019} derived an iterative formula to calculate the optical
conductivity of crystals to any order based on the knowledge of the Berry connections of the material.
Although the calculation of the linear optical conductivity of materials is straight forward and actually implemented
in most density functional calculation (DFT) software packages, the same cannot be said of the higher orders.
Given the technological importance of second order processes like second harmonic generation and photocondutivity,
or higher order like third harmonic generation, etc., the implementation of Ventura's formula departing from
first principles calculations like DFT seems to be very interesting.

But this endeavor faces a major hurdle.
The Berry connections in a crystal can be obtained by \cite{Resta1992,Zak1989}:
\begin{equation}\label{eq:berryconnection}
 \pmb{\xi}_{\pmb{k}ss'} = i\langle u_{\pmb{k}s}|\nabla_{\pmb{k}} u_{\pmb{k}s'} \rangle
                        = \frac{i}{v_{uc}}\int_{uc} d^3\pmb{r}\;u_{\pmb{k}s}^*(\pmb{r}) \nabla_{\pmb{k}} u_{\pmb{k}s'}(\pmb{r})
\end{equation}
which includes the gradient of the Bloch factor in reciprocal space.
In this equation, $s$ and $s'$ are electronic bands, $u_{\pmb{k}s}$ is the Bloch factor of the Bloch wavefunction at
reciprocal space point $\pmb{k}$ and band $s$, $v_{uc}$ is the volume of the unit cell and the integral is over the unit cell.
This means that we need the wavefunctions classified in bands where analyticity applies,
that is, are numerically differentiable in reciprocal space.
Strinati \cite{Strinati1978} has already noted that and proposed that bands should be labeled in such a way as to guarantee
the analyticity of the energy eigenvalues.

There are known problems in achieving this \cite{Strinati1978,Virk2007,Fukui2005}.
One is that the DFT programs calculate the Kohn-Sham equations for each $\pmb{k}$-point independently of the others.
Since the solutions of the differential equation come with an overall random phase,
that phase will be different from point to point, precluding analyticity.
The way this is dealt in the {\sc berry} package is explained in subsection \ref{ssec:randomphase}.
Another major obstacle is the degeneracy of the electronic bands at some points,
i.e. bands are not isolated but cross.
This problem also appears when trying to generate Wannier functions from DFT numerical calculations,
and it is dealt by isolating a set  of bands and performing a basis rotation on the set \cite{Marzari1997,Virk2007}.
Here the adopted procedure is similar, as can be seen in subsection \ref{ssec:basisrotation}.

Even surpassing these obstacles, the classification of the Bloch wavefunctions in analytic bands in an ordered way
is not simple.
{\sc berry} uses techniques of Graph Theory (GT) and Unsupervised Machine Learning (UML) iteratively to achieve the final result.
It also includes some visualization tools to help discern the quality of the calculation.

Berry connections and curvatures are then calculated and used to obtain
the linear and second harmonic generation optical conductivity.
Of course, once having the Berry connections, other non-linear optical properties can be implemented,
as well as other properties that can be calculated using the organization of wavefunctions in analytic bands.
We expect to implement them in the near future.

The paper is organized as follows.
In Sec. \ref{sec:algorithm} the algorithm and the implementation are described.
Then, in Sec. \ref{sec:usage} the instalation and usage of the software is explained.
In Sec. \ref{sec:examples} the software is applied to two materials: single layer hexagonal boron nitride and
single layer indium selenide, with some convergence studies.
For these two materials, we calculated linear and second harmonic conductivity
using the length gauge description of the reduced density matrix, as described below.
Other works on linear and non linear optical response of hBN using the length gauge description can
be found in  \cite{Hipolito2016} and \cite{Taghizadeh2017}.
Some considerations regarding the performance of {\sc berry} are exposed in Sec. \ref{sec:computational_profile}.
We comment on future perspectives of the software in Sec. \ref{sec:future} and
we draw some conclusions in Sec. \ref{sec:conclusions}.

\section{Algorithm and implementation}
\label{sec:algorithm}

\subsection{Preprocessing}
\label{ssec:preprocessing}

{\sc berry} departs from first principles calculations.
It aims at transforming the results obtained from a DFT software
(only {\sc Quantum Espresso} is currently supported,
but we expect to extend the possibility to other DFT software)
in such a way as to be able to use in other calculations.

So the first step is to obtain the results from a DFT calculation.
This is accomplished by a package that,
given a self consistent field (scf) input file of {\sc Quantum Espresso} (QE),
with the pseudopotentials for the DFT run,
and a small input file with the area and density of points in the reciprocal space wanted,
automatically runs the DFT scf calculation,
generates the non-scf (nscf) input file for the $\pmb{k}$-points wanted and runs it.
This procedure assures that the wavefunctions are well ordered
and with their position identified in reciprocal space.

The package also reads all relevant data from the DFT run
and saves them in a manner useful for the other packages of the software.
It works like an interface between the DFT software and the rest of the packages,
and so is adaptable for use with other DFT packages other than QE.

\begin{figure}[h]
  \centering
  \includegraphics[scale=0.8,keepaspectratio=true]{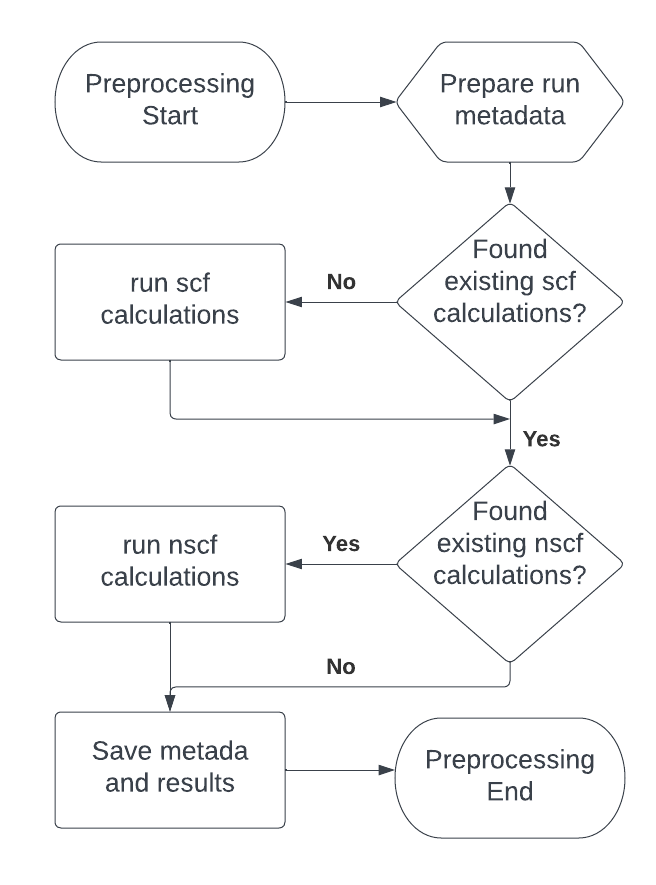}
  \caption{Flux chart for the preprocessing implementation.
  This part of the code runs the DFT calculations and extracts the data for the run, saving it for later use.}
  \label{fig:flowchart_preprocess}
\end{figure}

\subsection{Eliminating random phases}
\label{ssec:randomphase}

The next step is to read the wavefunctions obtained from the DFT calculation,
make them coherent by multiplying by a properly chosen phase,
and save them in a format suitable for the next steps.

Wavefunctions in a crystaline material can be written as:
\begin{equation}
 \psi_{\pmb{k}s}(\pmb{r}) = e^{i\pmb{k}\cdot\pmb{r}}u_{\pmb{k}s}(\pmb{r})
\end{equation}
with the Bloch factor $u_{\pmb{k}s}(\pmb{r}) = u_{\pmb{k}s}(\pmb{r} + \pmb{R})$
for $\pmb{R}$ any lattice vector.

A procedure to solve this problem is hinted, for instance, in \cite{Ferreira1970} and \cite{Pedersen1991}.
The idea is that all wavefunctions have to have a common reference.
We choose that the phases of all wavefunctions of the system should be synchronized in a chosen point $\pmb{r}_0$ of real space.
For instance, we can set the phase to zero at that chosen point by
 multiplying each wavefunction $\psi_{\pmb{k}s}(\pmb{r})$
 by the reciprocal of the phase $e^{i\phi_0}$ it has in that point:
$\psi_{\pmb{k}s}(\pmb{r}) \rightarrow e^{-i\phi_0}\psi_{\pmb{k}s}(\pmb{r})$,
where $\phi_0$ is the argument of wavefunction $\psi_{\pmb{k}s}(\pmb{r})$ at $\pmb{r} = \pmb{r}_0$.
The choice of the point in real space is not relevant
(as long as all wavefunctions are different from zero at that point)
 since choosing another point would simply add a global phase to all wavefunctions,
 that would cancel in equation \ref{eq:berryconnection}.
 A low symmetry point in real space is chosen so that
 wavefunctions at that point in real space are different from zero,
 a condition that has to be verified.
This results in a totally coherent set of wavefunctions, for which we can apply the usual operations.

Wavefunctions are saved in individual files, labeled by $\pmb{k}$-point number and band,
and are written as a function of position, already corrected for the phase.

\subsection{Establishing the bands: Solver Algorithm}
\label{ssec:cluster}

This is the main part of the project,
by which each eigenstate is put in a set that form a continuous and analytic band.
By analytic band we mean that the gradient of the Bloch factors
$\pmb{\nabla}_{\pmb{k}}u_{\pmb{k}s}$ of the set $s$ that we call a band
may be calculated.

There are two main data that can be used to establish to which band a eigenstate belongs.
One is the eigenenergy $E_{\pmb{k},s}$, which has to be continuous in $\pmb{k}$ in every band $s$.
This criterion is necessary but clearly insufficient, since there are degenerate points
(i.e. with the same energy, although belonging to different bands)
and even if the states are not degenerate they can be sufficiently close in energy to be difficult
to discern numerically if there is continuity or not.

A more natural criterium is the continuity of the Bloch factors $u_{\pmb{k}s}$ along the reciprocal space,
which are the ones used to calculate the gradient.
Continuity means that $u_{\pmb{k}s}$ should change little from one $\pmb{k}$ to another very close $\pmb{k}'$.
Therefore the dot product of two neighboring Bloch factors $u_{\pmb{k}s}$ and $u_{\pmb{k}'s}$ should be close to $1$
(they are normalized)
if there is continuity between the factors in the same band $s$.
Otherwise, the dot product should be close to zero, since Block factors of eigenstates in the same
$\pmb{k}$-point are orthogonal.

To verify the continuity of the Bloch factors we calculated the inner product
$S_{\pmb{k}\pmb{k}'ss'} = \langle u_{\pmb{k}s}(\pmb{r})|u_{\pmb{k}'s'}(\pmb{r})\rangle$ between
Bloch factors of neighboring $\pmb{k}$ and $\pmb{k}'=\pmb{k}+\Delta\pmb{k}$ points.

For sufficiently small $\Delta\pmb{k}$ the modulus of the inner product of the normalized Bloch factors $|S|$
 is either close to one for states that vary continuously and belong to the same band
 or much smaller than one if the states belong to different bands.

This way we were able to classify, for each $\pmb{k}$, which eigenstate corresponds (in the sense that it is continuous)
to which eigenstate of neighbor $\pmb{k}+\Delta\pmb{k}$ point.
We then propagate this procedure through all $\pmb{k}$ points of the Brillouin Zone (BZ) and all eigenstates
and make sets of eigenstates that are linked by continuity.
This gives a straight forward way of determining which solutions $|u_{\pmb{k}s}\rangle$ should go together to form a band.

Unfortunately, although this works for most of the cases, there are some for which it doesn't.
One example where it doesn't work is where we have the degenerate states,
since any linear combination of degenerate states is also a solution to the eigenvalue problem;
thus the orthogonality criterion fails.

To address the cases where these criteria fail, we resort to Graph Theory (GT)
and Unsupervised Machine Learning (UML).
Our approach involves two steps, that are repeated iteratively, and is called \emph{Solver Algorithm}.
The first step is a component identification based on GT and the second is
a samples assignment using UML methods.

The \emph{vertices} or \emph{nodes} of GT are points of the form $p=(\pmb{k},E_{\pmb{k},s})$
and a \emph{connection} or \emph{edge} between points is formed whenever
$|S_{\pmb{k}\pmb{k}'ss'}| > \textrm{TOL}$,
where $\textrm{TOL}$ is a tolerance the user can define.
For simplicity, this graph is undirected and does not have weights on its edges.



Component detection algorithms based on Graph Theory are used to identify groups of points for which the dot product of the Bloch factors $S_{\pmb{k}\pmb{k}'ss'}$ is close to 1.
These groups are referred to as \emph{components} \cite{newman2018networks}. Specifically, we use the algorithm implemented by the "Networkx" Python library \cite{SciPyProceedings_11}.
Once identified, these components correspond to patches of a continuous band. These patches have to be "glued" together consistently to form the full band.


This prepares an initial data structure in the form of components, convenient for an Unsupervised Machine-Learning clustering algorithm.
The purpose of UML algorithms is to identify patterns within unclassified data by analyzing the differences or similarities between data points using specific metrics.
One of the most commonly used UML algorithms is the K-means clustering algorithm. This algorithm classifies the data points into a predefined number of \emph{clusters} refered to a collection of data points that exhibit similarities according to a specific metric \cite{Alloghani2020,UnsupervisedSinaga}.

The implemented UML algorithm in this program is based on the K-means approach, where the component detection method is used to identify the initial K clusters and data points to be classified. However, the current implementation does not rely on pre-existing algorithms, as some data points may require adjustments to be correctly assigned.

The implementation of the algorithm offers an added advantage, as it provides the flexibility to adjust and correctly assign data points for more accurate classification.

\begin{figure}[h]
  \centering
 \includegraphics[scale=0.6,keepaspectratio=true]{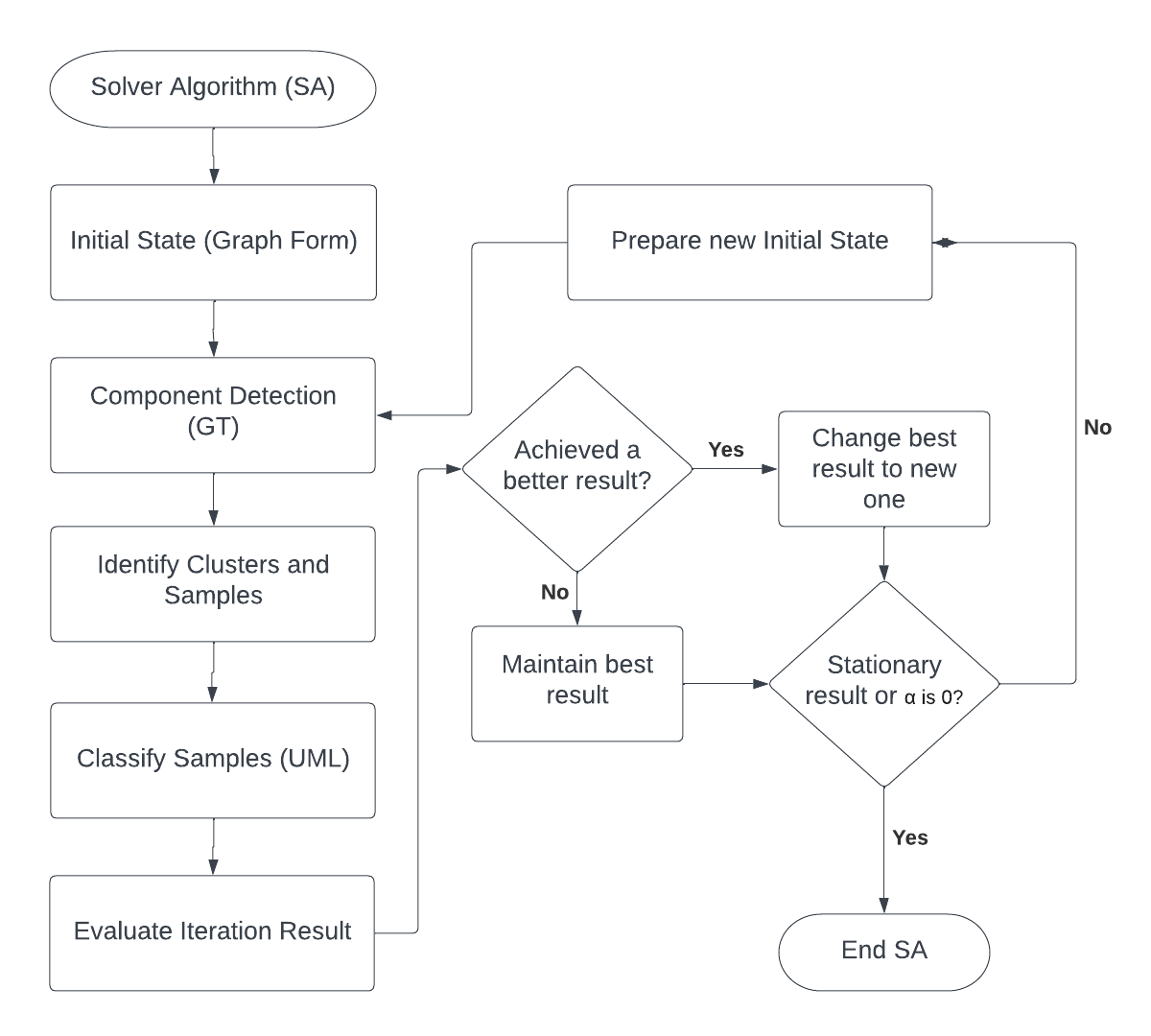}
 \caption{The flux chart represents a simplified Solver Algorithm version.
 It starts converting the problem into a graph form using the dot-product data.
 The graph’s components are detected by Graph Theory (GT) methods.
 Then, the clusters and samples are identified to be classified with Unsupervised Machine Learning (UML) techniques.
 A result evaluation identifies the initial state for the next iteration.
 The algorithm stops when it reaches a stationary result, or the relaxation parameter $\alpha$ is zero.}
 \label{fig:flowchart_SA}
\end{figure}

UML then forms groups of components called \emph{clusters}
(which are patches of bands glued together).
Each  data point belongs exclusively to a single cluster.
Ideally, in the end, each cluster should contain all points of a continuous
(in the analytic sense) band and only those.
Yet this usually is not enough to build complete bands,
and so the process has to be iterated,
as shown in figure \ref{fig:flowchart_SA}.
Each iteration uses as a starting point the best result of previous iterations.
One iteration involves identifying the components, given the initial state in the form of a graph.
Clustering these components gives larger components,
that will form a new state that will be used as a initial state of the next iteration.
The optimization process ends when the result does not change between iterations,
reaching a stationary solution, or the relaxation parameter reaches the minimum value.

The final result is evaluated to ascertain which bands are correctly built
and for the identification of problems such as degenerate states,
that have to be dealt afterwards.
A table is then produced with this information
and the number of useful bands generated is given,
for the next packages of the software.

\subsection{Dealing with degeneracy}
\label{ssec:basisrotation}

Degenerate points are easy to spot, since they have the same energy, at the same $\pmb{k}$-point.

Yet, there is no guarantee that the Bloch factors of these degenerate states are continuous
to the adjacent Bloch factors, that are already attributed to specific bands,
since a linear combination of two degenerate eigenstates is also an eigenstate.

The solution is to make a local basis rotation of these two states,
by applying an unitary transformation in such a way as to maximize
the continuity with the two bands that cross.
This is what is done for similar problems when constructing Wannier functions.
It can be applied after the previous step of establishing the bands,
to the identified degenerate points.

 Consider a point in $\pmb{k}$-space and two Bloch factors $|u_{\pmb{k}s}\rangle$ and $|u_{\pmb{k}s'}\rangle$
 that have been signaled as not continuous to the neighboring Bloch factors $|u_{\pmb{k}'A}\rangle$ and $|u_{\pmb{k}'B}\rangle$
 that belong to band $A$ and band $B$, respectively.

 We will apply an unitary transformation
 \begin{align}
  |u_{\pmb{k}A}'\rangle &= a_1|u_{\pmb{k}s}\rangle + a_2|u_{\pmb{k}s'}\rangle \\
  |u_{\pmb{k}B}'\rangle &= b_1|u_{\pmb{k}s}\rangle + b_2|u_{\pmb{k}s'}\rangle
 \end{align}
 where we have the restrictions due to orthonormalization:
 \begin{align}
  a_1a_1^* + a_2a_2^* &= 1\\
  b_1b_1^* + b_2b_2^* &= 1\\
  a_1^*b_1 + a_2^*b_2 &= 0
 \end{align}

 Then we want this transformation to be such that maximizes continuity to bands $A$ and $B$, so that
 the dot products $\langle u_{\pmb{k}'A}|u_{\pmb{k}A}'\rangle $ and $\langle u_{\pmb{k}'B}|u_{\pmb{k}B}'\rangle $ are close to $1$,
 which is the maximum they can be.

 We look for the set $a_1, a_2, b_1, b_2$ that give:
 \begin{align}
  \max \langle u_{\pmb{k}'A}|u_{\pmb{k}A}'\rangle &=
  \max \left( \langle u_{\pmb{k}'A}|u_{\pmb{k}s}\rangle a_1 + \langle u_{\pmb{k}'A}|u_{\pmb{k}s'}\rangle a_2\right) \\
  \max \langle u_{\pmb{k}'B}|u_{\pmb{k}B}'\rangle &=
  \max \left( \langle u_{\pmb{k}'B}|u_{\pmb{k}s}\rangle b_1 + \langle u_{\pmb{k}'B}|u_{\pmb{k}s'}\rangle b_2\right)
 \end{align}

 Using the Bloch factors in the rotated basis assures the analyticity at the degenerate point.

\subsection{Calculating Berry connections}
\label{ssec:gradient}

Having obtained the set of states that belong to each band of interest,
ordered in reciprocal space and covering all BZ,
we can then determine the Berry connections between every pair of bands $s$ and $s'$
at each $\pmb{k}$ point of the BZ.

To calculate the gradient in $\pmb{k}$-space, $\nabla_{\pmb{k}} u_{\pmb{k}s}$, we have to have
the Bloch factors written as a function of $\pmb{r}$ for each $\pmb{k}$-point and not as a function of $\pmb{k}$
for each real space point $\pmb{r}$.
So a conversion is done before differentiation.

Gradients are then calculated and saved in files.
Berry connections are calculated for each pair of bands using Equation \ref{eq:berryconnection}, and saved in files.

Berry curvatures can also be calculated using Equation \ref{eq:curvature} \cite{Gradhand_2012}
\begin{equation}\label{eq:curvature}
 \pmb{\Omega}_{s,s'} = i\int_{uc}d^3\pmb{r}\;(\nabla_{\pmb{k}}u_{\pmb{k}s}^* \times \nabla_{\pmb{k}} u_{\pmb{k}s'})
\end{equation}

\subsection{Optical conductivity}
\label{ssec:opticalcondutivity}

Now it is possible to proceed to the calculation of the linear optical conductivity and the second harmonic generation.
For this, we use the length gauge description of the reduced density matrix \cite{Sipe1995,Sipe2000,Ventura2017},
as described in reference \cite{Ventura2017}.

For the case of a cold 2 dimensional semiconductor, the linear conductivity can be expressed in terms of a single, interband, contribution,
 given by Equation \ref{eq:first_order}
\begin{equation}\label{eq:first_order}
 \sigma^{\beta\alpha}(\omega) = \frac{i e^2}{\hslash}\sum_{s's}\int\limits_{BZ}\frac{d^2\pmb{k}}{(2\pi)^2}
 \frac{\Delta\epsilon_{\pmb{k}s's} \xi_{\pmb{k}ss'}^{\beta}\xi_{\pmb{k}s's}^{\alpha}}{(\hslash\omega + i\Gamma)
 - \Delta\epsilon_{\pmb{k}s's}}\Delta f_{\pmb{k}ss'},
\end{equation}
where $\Delta\epsilon_{\pmb{k}s's} = \epsilon_{\pmb{k}s'} - \epsilon_{\pmb{k}s}$,
 $\Delta f_{\pmb{k}ss'} = f_{\pmb{k}s} - f_{\pmb{k}s'}$,
 $\epsilon_{\pmb{k}s}$ is the eigenvalue of band $s$ at $\pmb{k}$, and $f_{\pmb{k}s}=f(\epsilon_{\pmb{k}s})$ is the
Fermi-Dirac distribution function.
Note that with the knowledge of the band energies, $\epsilon_{\pmb{k}s}$,
and the Berry connections, $\xi_{\pmb{k}ss'}^{\beta}$, both of which can be determined by a DFT calculation,
we can compute the optical response of a crystal.
The indexes $\alpha$ and $\beta$ run through the spatial coordinates $x$ and $y$
as we are dealing with a two dimensional material, and the conductivity tensor has, in general, four relevant components.
The results from this calculation (see Section \ref{sec:examples}) are the same as the ones obtained
on the same system calculated using the dipole approximation within
the random phase approximation (RPA) \cite{Brener1975}, that usually is available in DFT packages.
This very good agreement validates the model:
 the same physical quantity calculated in two totally different models gives the same result.

The second order response can be calculated for the case of the second harmonic generation.
Following the procedure described in ref. \cite{Ventura2017}
we obtain, for a diagonal component of the SHG, $\sigma^{\beta\beta\beta}(\omega,\omega)$, Eq.(\ref{eq:second_order}).
\begin{eqnarray}\label{eq:second_order}
 & \nonumber\sigma^{\beta\beta\beta}(\omega,\omega) = -\dfrac{e^3}{\hslash}\mathlarger{\int}\limits_{BZ}\dfrac{d^2\pmb{k}}{(2\pi)^2}
 \left[
 \mathlarger{\sum}\limits_{s'\ne s}\dfrac{1}{2} \dfrac{\Delta\epsilon_{\pmb{k}ss'} }{ g_{\omega\pmb{k}ss'}}
 \bigg\lbrace
 \left[ \xi_{\pmb{k}s's}^{\beta},(\xi_{\pmb{k}ss'}^{\beta})_{;\beta}\right]
 \dfrac{\Delta f_{\pmb{k}s's}}{  h_{\omega\pmb{k}ss'} }
 \bigg\rbrace
 \right.\\
 & -\dfrac{i}{2}\mathlarger{\sum}\limits_{s'\ne s \ne r}
 \dfrac{\Delta \epsilon_{\pmb{k}ss'} }{ g_{\omega\pmb{k}ss'}}
 \left
  \lbrace \dfrac{\Delta f_{\pmb{k}s'r}}{h_{\omega\pmb{k}rs'}} - \dfrac{\Delta f_{\pmb{k}rs }}{h_{\omega\pmb{k}sr}} \bigg\rbrace
 \left( \xi_{\pmb{k}s's}^{\beta} \xi_{\pmb{k}sr}^{\beta} \xi_{\pmb{k}rs'}^{\beta}
     +  \xi_{\pmb{k}s'r}^{\beta} \xi_{\pmb{k}rs}^{\beta} \xi_{\pmb{k}ss'}^{\beta}\right)
 \right]
\end{eqnarray}
where
\begin{equation}
 g_{\omega\pmb{k}ss'} = 2(\hslash\omega + i\Gamma) - \Delta\epsilon_{\pmb{k}ss'}
\end{equation}
\begin{equation}
 h_{\omega\pmb{k}ss'} = (\hslash\omega + i\Gamma) - \Delta\epsilon_{\pmb{k}ss'}
\end{equation}
\begin{equation}
  \left[ \xi_{\pmb{k}s's}^{\beta},(\xi_{\pmb{k}ss'}^{\beta})_{;\beta}\right] =
  \xi_{\pmb{k}s's}^{\beta}(\xi_{\pmb{k}ss'}^{\beta})_{;\beta} - (\xi_{\pmb{k}ss'}^{\beta})_{;\beta}\xi_{\pmb{k}ss'}^{\beta}
\end{equation}
\begin{equation}
 (\xi_{\pmb{k}ss'}^{\beta})_{;\beta}=
  \left( \nabla^{\beta} \xi_{\pmb{k}ss'}^{\beta}  \right) - i\left( \xi_{\pmb{k}ss}^{\beta} - \xi_{\pmb{k}s's'}^{\beta}\right) \xi_{\pmb{k}ss'}^{\beta}
\end{equation}

\section{Installation and Usage}
\label{sec:usage}

\subsection{Installation}
\label{ssec:instalation}

The {\sc berry} software can be installed using pip package manager with the command
 \begin{verbatim}
  pip install berry-suite
 \end{verbatim}
and all the dependencies will be installed automatically.
To install the software with the development dependencies, use:
 \begin{verbatim}
  pip install berry-suite[dev]
 \end{verbatim}

Alternatively, the software can be downloaded from the GitHub repository,
and installed using the pip command
 \begin{verbatim}
  pip install -e .
 \end{verbatim}
A working {\sc Quantum ESPRESSO} installation, version v.6.6 or higher, is required.
{\sc berry} was tested with python 3.8 and 3.10, but likely works with other versions of python3.

\subsection{Usage}
\label{ssec:usage}

The {\sc berry} software has a Command Line Interface (CLI), that is used to run the packages.
 \begin{verbatim}
  berry [package options] package parameter [package options]
 \end{verbatim}
where the possible packages are shown in Table \ref{tab:packages}
and the parameter is dependent on which package is being run.
Simply writing \verb|berry| will output the help message. From there, the available options can be seen.

\begin{table}[htp]
 \centering
  \caption{List of packages included in the software.}
 \begin{tabular}{lr}
 \hline
 Package & Description \\
 \hline
 \verb|preprocess| & Runs DFT calculations and prepares for the following runs \\
 \verb|wfcgen| & Reads wavefunctions from DFT results and saves them \\
 \verb|dot| & Calculates the dot product of the Bloch factor \\
 \verb|cluster| & Finds the bands \\
 \verb|basis| & Rotates basis of degenerate states \\
 \verb|r2k| & Converts from $\pmb{r}$-space to $\pmb{k}$-space \\
 \verb|geometry| & Calculates Berry connections and curvatures \\
 \verb|condutivity| & Calculates linear optical conductivity \\
 \verb|shg| & Calculates SHG optical conductivity \\
 \hline
 \end{tabular}
 \label{tab:packages}
\end{table}

The \verb|preprocess| package is the first one to be run.
A working directory has to be created, with a directory called \verb|dft| in it,
with the files needed for running QE for the desired system.
It needs an input file (given as a parameter) with the number of $\pmb{k}$-points in each direction
and the area in reciprocal space that is covered.
It can include some change to the default values of the run.

The rest of the packages should be run in the order given in Table \ref{tab:packages}
and no input is needed except after the \verb|cluster| package,
where the number of the highest band has to be given.
This is the highest analytic band that was possible to recover,
and can be read from the output of the \verb|cluster| package.

All packages give information about the run in a file named after the name of the package with a \verb|.log| extension.
Also, help can be obtained adding \verb|-h| as an option to the package.

\subsection{Parallelization}
\label{ssec:parallelization}



{\sc berry} supports parallel execution of QE tasks on a single or multiple interconnected multicore servers.
It relies on QE's MPI-based multiprocess parallelization \cite{qe_parallelization}
through the configuration of specific tasks based on user input.
The number of processes can be specified when running any of {\sc berry}'s packages through an input file
in the case of \verb|preprocess| and \verb|wfcgen|, or by the \verb|-np| CLI argument for the rest of the packages.

While this is initially the best approach to parallelize {\sc berry},
since it requires a smaller effort for significant gains in performance
and ensures that it benefits from future improvements in QE, it may be limiting in some edge cases.
Section \ref{sec:computational_profile} analyses the scalability limitations of its parallelization,
which could be further improved in the future.

\subsection{Output visualization}
\label{ssec:visualization}

A set of packages was created for a quick visualization of the results and debugging.
To run them, one has to use the command:
\begin{verbatim}
 berry-vis [-h] package options
\end{verbatim}
where the package can be \verb|geometry|, \verb|debug| or \verb|wave|.
Table \ref{tab:packages_vis} gives a summary of the possibilities.

\begin{table}[htp]
 \centering
  \caption{List of packages for visualization.}
 \begin{tabular}{lr}
 \hline
  Package  & Description \\
 \hline
  debug    &   Prints data for debugging \\
  geometry &   Draws Berry connection and curvature \\
  wave     &   Shows the electronic band structure \\
 \hline
 \end{tabular}
 \label{tab:packages_vis}
\end{table}

Each program contains a set a sub-commands that can be used to visualize different aspects of the results.
The help message of each package gives a summary of the available options.

\section{Examples}
\label{sec:examples}

The software was tested for two materials, namely single layer hexagonal boron nitride (hBN) and InSe.
For both materials, the software was ran using different meshes of points in reciprocal space
and different energy cutoffs for the DFT runs.

The higher the energy cutoff, the higher the numerical precision of the calculation,
since a larger energy cutoff implies a larger base of plane waves to describe the wavefunctions.
The numerical precision increases monotonically with the size of the plane wave basis,
but also the size of the files that store the wavefunctions.

A higher density of points in reciprocal space allows for a better construction of the analytic band,
and a better precision in the calculation of the gradient of the wavefunctions.
But the number of wavefunctions is proportional to the number of $\pmb{k}$-points,
and so the amount of calculus needed.
Convergence tests are then convenient.

The DFT calculations were performed using {\sc Quantum ESPRESSO}~\cite{Giannozzi2009,Giannozzi2017}.
The exchange-correlation functional used was the generalized gradient approximation
of Perdew-Burke-Ernzerhof (GGA-PBE)\cite{Perdew1996}
and the integration over the Brillouin-zone was performed using the scheme proposed by
Monkhorst-Pack~\cite{Monkhorst1976}  with a grid of $24\times 24 \times 1$ $\pmb{k}$-points.
A vacuum size of 30 bohr was used to avoid interactions between the periodic images.
$\Gamma$, the relaxation parameter, is fixed to
$\Gamma=0.01$~Ry $=0.136$~eV in all calculations.

\subsection{hBN}
\label{ssec:hBN}

For hBN scalar-relativistic norm-conserving pseudopotentials \cite{Hamann2013,Schlipf2015} were used.
The energy cutoff and the density of $\pmb{k}$-points were varied
(the area covered in reciprocal space is always the Brillouin Zone (BZ) or equivalent)
with different results for the number of correct bands obtained, as shown in Table \ref{tab:hBN}.
It is clear that the density of $\pmb{k}$-points is important in order to achieve a larger number of
successful reconstructed bands.
On the other hand, there is no such improvement with energy cutoff, perhaps because the values used
were already very high.
Also, there is some inconsistency for the lower number of $\pmb{k}$-points
where increasing the energy cutoff doesn't necessarily leads to improvement.
Nevertheless, detailed examination of the output of the \verb|cluster| package
shows that for $E_{cut}=120$~Ry the 8th band is correct and there is only one
$\pmb{k}$-point of the 7th band that is estimated to be wrong, and so we decided to proceed with eight bands.
The same happened for the 9th and 10th bands of $E_{cut}=80$~Ry for the larger density of $\pmb{k}$-points,
and so we proceeded with 10 bands.

\begin{table}[htp]
 \centering
  \caption{Number of correct bands achieved for different values of the energy cutoff $E_{cut}$ and density of $\pmb{k}$-points.
  In parentesis is the number actually used for optical conductivity calculation.}
 \begin{center}
\begin{tabular}{lcr}
 \hline
 $\pmb{k}$-points & $E_{cut}$ (Ry) & Bands \\
 \hline
 $47\times41$ &  80 & 6 \\
 $47\times41$ & 100 & 8 \\
 $47\times41$ & 120 & 6 (8) \\
 $47\times41$ & 150 & 8 \\
 \hline
 $60\times52$ &  80 & 8 (10) \\
 $60\times52$ & 100 & 10 \\
 $60\times52$ & 120 & 10 \\
 $60\times52$ & 150 & 10 \\
 \hline
 \end{tabular}
 \end{center}
 \label{tab:hBN}
\end{table}

\begin{figure}[ht]
 \centering
 \includegraphics[scale=0.5,keepaspectratio=true]{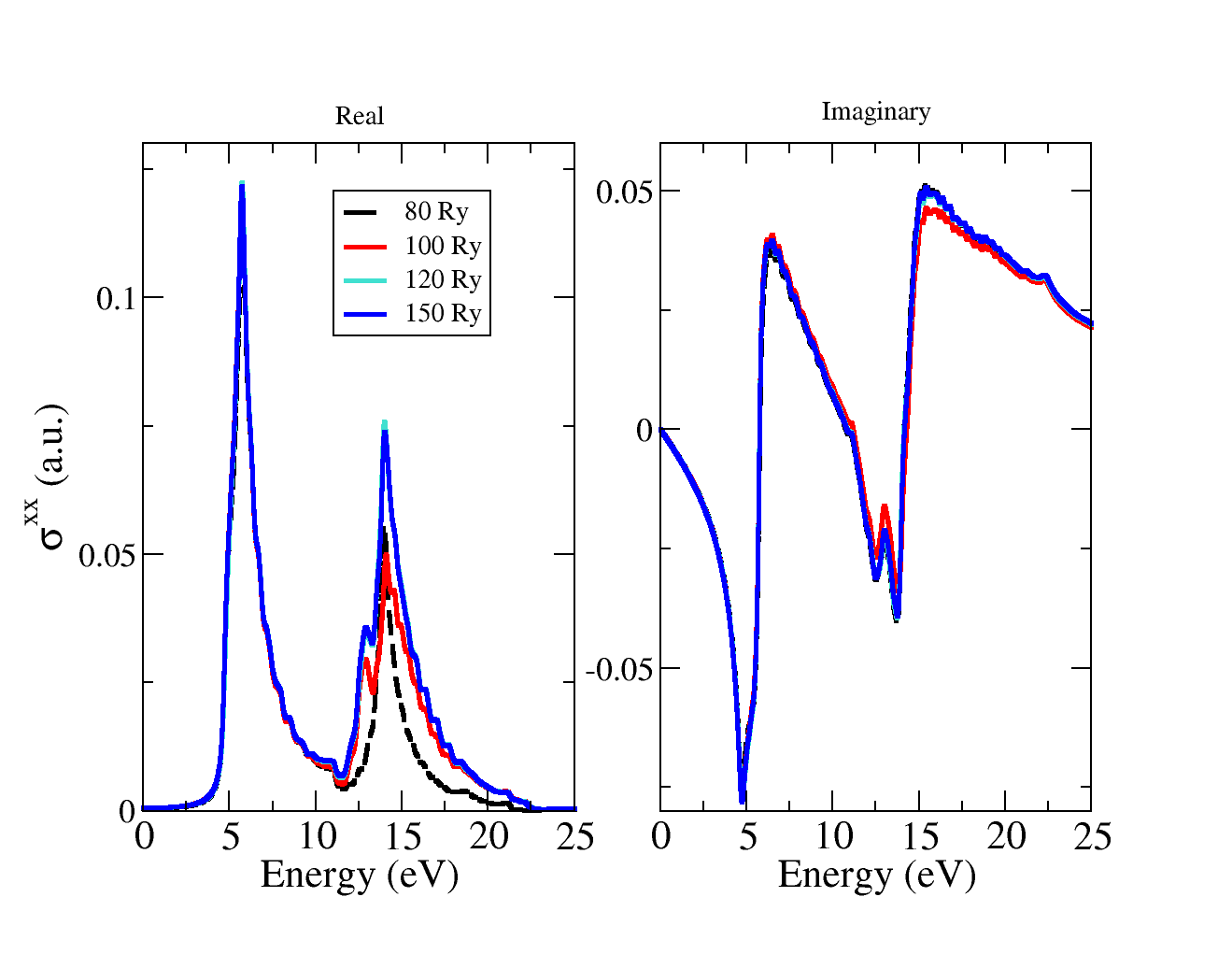}
 \caption{Linear optical condutivity of hBN for several energy cutoffs, and $47\times41$ $\pmb{k}$-points.}
 \label{fig:condutivity1927}
\end{figure}

\begin{figure}[ht]
 \centering
 \includegraphics[scale=0.5,keepaspectratio=true]{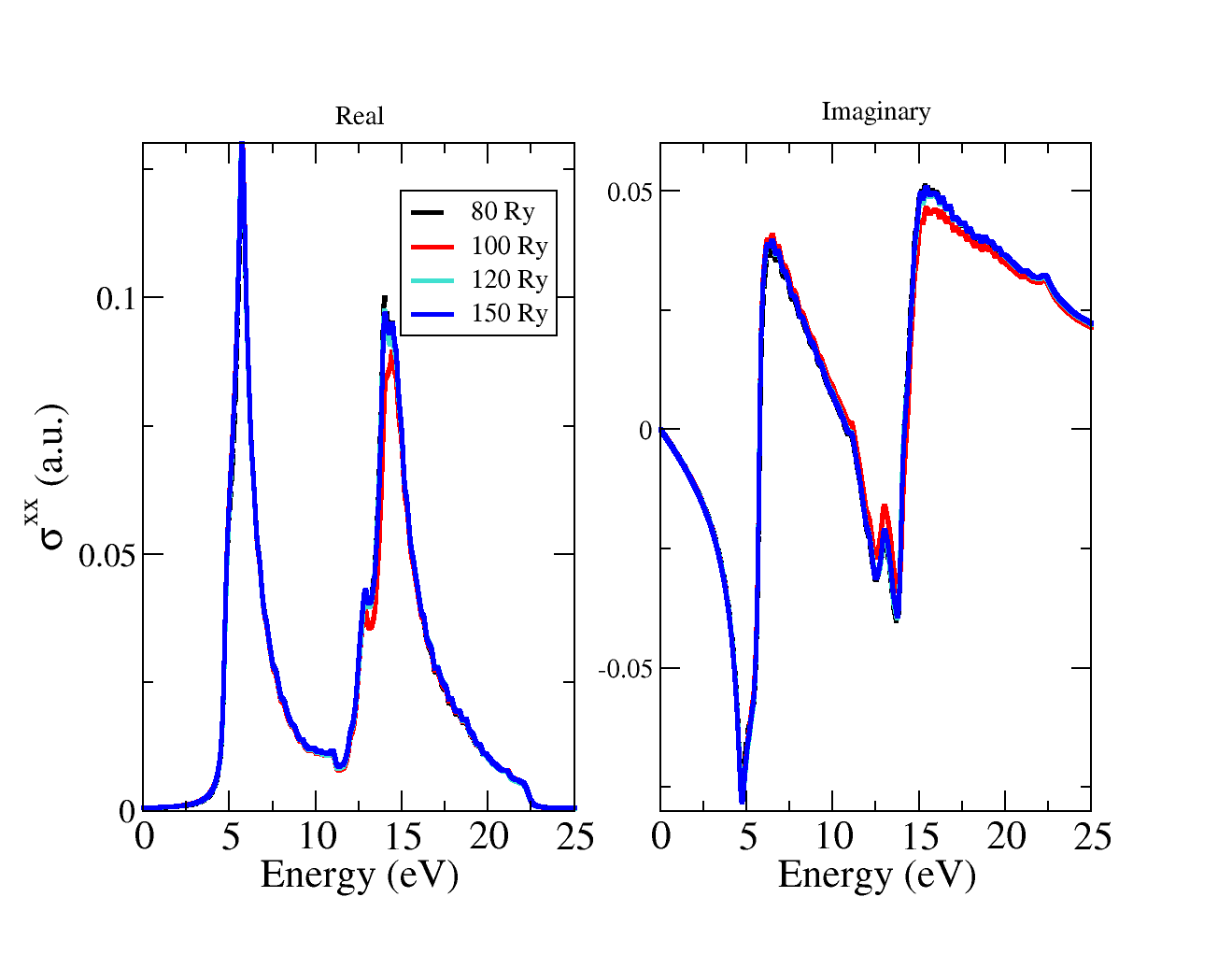}
 \caption{Linear optical condutivity of hBN for several energy cutoffs, and $60\times52$ $\pmb{k}$-points.}
 \label{fig:condutivity3120}
\end{figure}

Figure \ref{fig:condutivity1927} shows the real and imaginary parts of the optical condutivity for $47\times41$ $\pmb{k}$-points,
while Figure \ref{fig:condutivity3120} shows the same for $60\times52$ $\pmb{k}$-points.
In both cases, the dependence in energy cutoff is minimal. The greatest difference is for $E_{cut}=80$~Ry and $47\times41$ $\pmb{k}$-points
but that is because we are using just 6 bands, while the others use 8, as shown in Table \ref{tab:hBN}.

 In this case, the number of bands afects the height and ratio of the main peaks, but that is natural,
and is the only relevant difference between the results shown in Figure \ref{fig:condutivity1927} and Figure \ref{fig:condutivity3120}.

\begin{figure}[ht]
 \centering
 \includegraphics[scale=0.25,keepaspectratio=true]{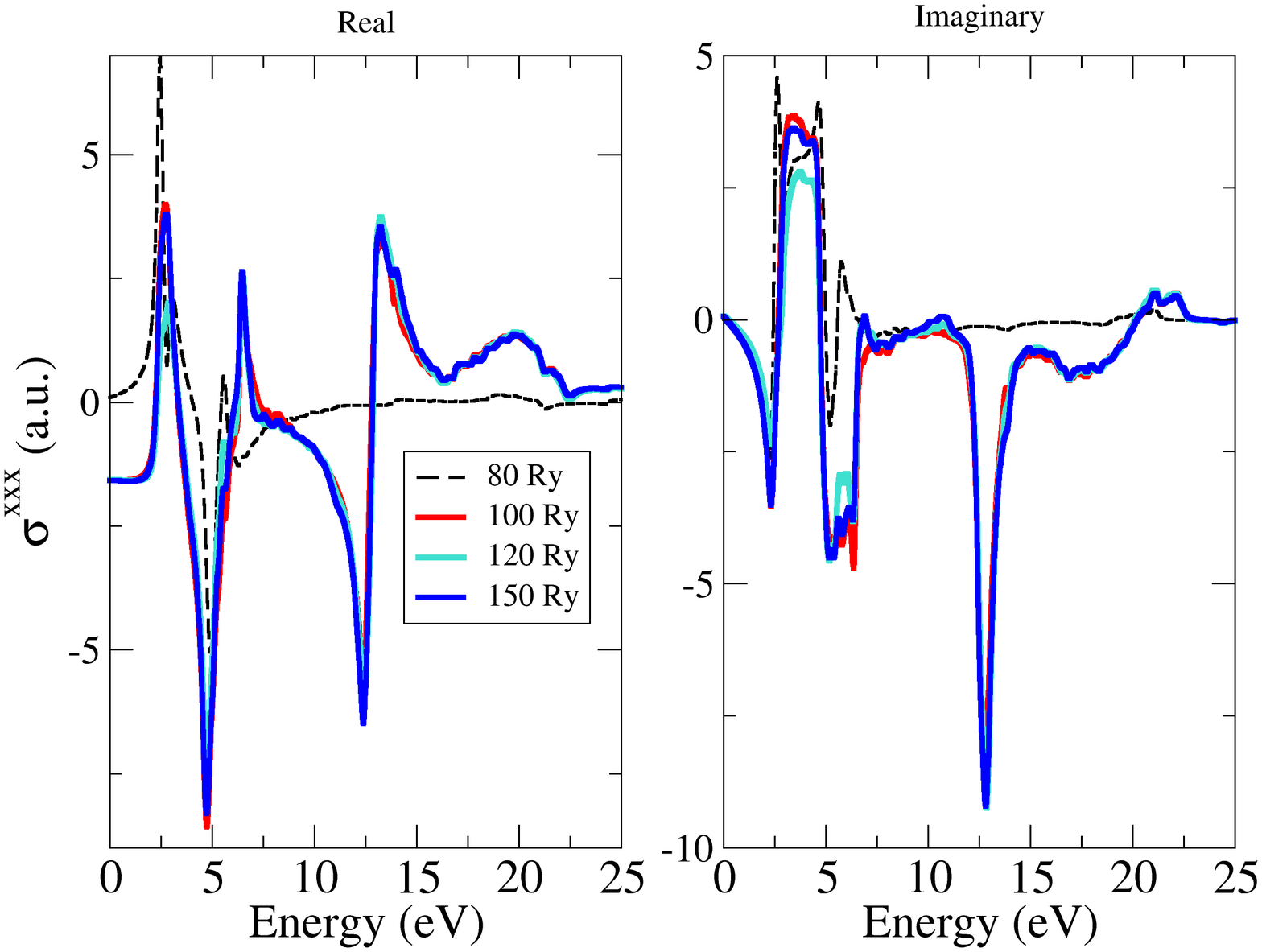}
 \caption{Second harmonic generation optical conductivity of hBN for several energy cutoffs, and $47\times41$ $\pmb{k}$-points.}
 \label{fig:shg1927}
\end{figure}

\begin{figure}[ht]
 \centering
 \includegraphics[scale=0.25,keepaspectratio=true]{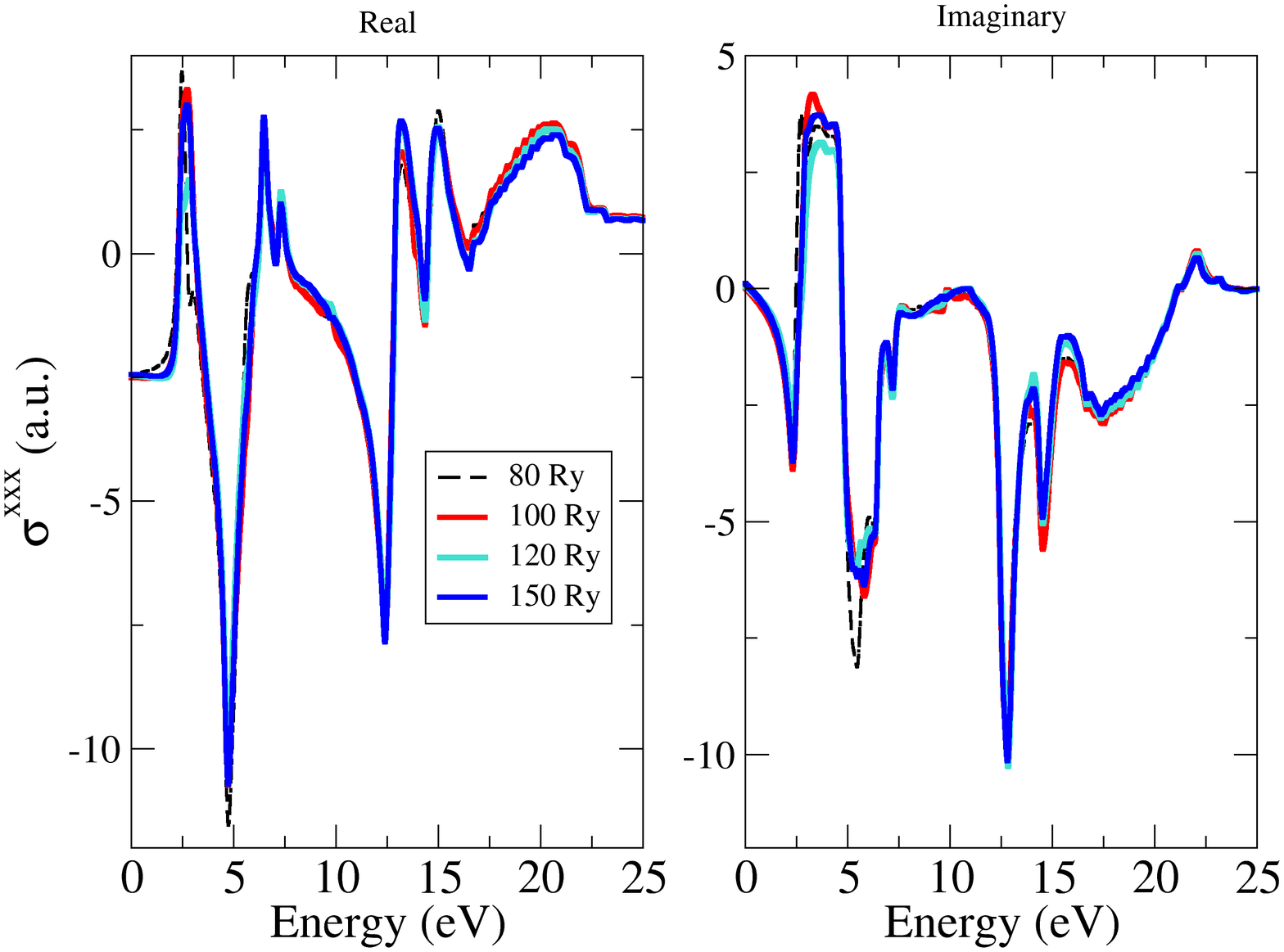}
 \caption{Second harmonic generation optical conductivity of hBN for several energy cutoffs, and $60\times52$ $\pmb{k}$-points.}
 \label{fig:shg3120}
\end{figure}

The Second Harmonic Generation optical conductivity is shown in Figures \ref{fig:shg1927} and \ref{fig:shg3120}.
The same behaviour as for optical conductivity can be seen:
the number of bands that were used is what actually makes a difference.

The dependence on energy cutoff is small (only noticible for the lowest value of 80 Ry)
and the number of $\pmb{k}$-points used
seems to affect the result indirectly via the number of bands that can be solved
rather than directly changing the result.
The position of the peaks does not change
but their intensity increases with the number of bands,
and for higher energies more peaks appear, as expected,
since there are more bands with higher energy.

\subsection{InSe}
\label{ssec:InSe}

For InSe full relativistic norm-conserving pseudopotentials
with nonlinear core-correction and spin-orbit information were used.
Table \ref{tab:InSe} shows the number of correct bands achieved for two different sets of $\pmb{k}$-points and two energies.
There are 18 electrons in InSe, so 9 bands are full.
For $47\times41$ $\pmb{k}$-points only two conduction bands were obtained which is a low number.
Increasing the number of $\pmb{k}$-points to $59\times51$ improved the conduction bands achieved to seven.
Yet, energy cutoff lower than 100 Ry lead to less than 9 solved bands,
which is useless for the optical condutivity calculations.

\begin{table}[htp]
 \centering
  \caption{Number of correct InSe bands achieved for different values of the energy cutoff $E_{cut}$ and density of $\pmb{k}$-points.}
 \begin{center}
\begin{tabular}{lcr}
 \hline
 $\pmb{k}$-points & $E_{cut}$ (Ry) & Bands \\
 \hline
 $47\times41$ & 100 & 11 \\
 $47\times41$ & 120 & 11 \\
 \hline
 $59\times51$ & 100 & 16\\
 $59\times51$ & 120 & 16 \\
 $59\times51$ & 150 & 16 \\
 \hline
 \end{tabular}
 \end{center}
 \label{tab:InSe}
\end{table}

\begin{figure}[h]
 \centering
 \includegraphics[scale=0.25]{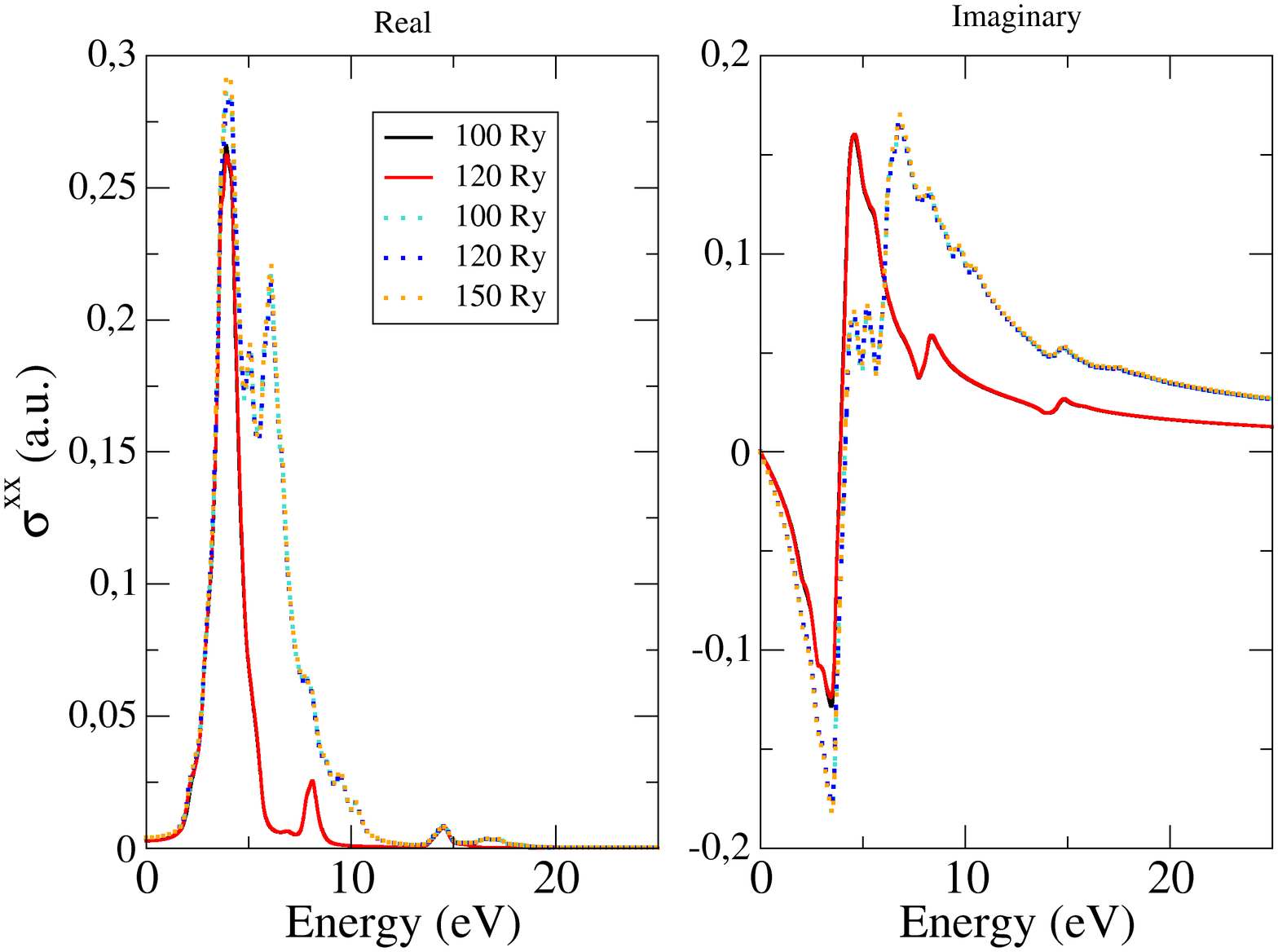}
 \caption{Linear optical condutivity of InSe for different cutoff energies and density of $\pmb{k}$-points.
 Lines are for $47\times41$ $\pmb{k}$-points and dots are for $59\times51$ $\pmb{k}$-points.}
 \label{fig:condutivity_inse}
\end{figure}

Figure \ref{fig:condutivity_inse} shows the linear optical conductivity of InSe for the five cases shown in Table \ref{tab:InSe}.
Naturally, there is a notable difference between the cases where only 11 bands were considered and the ones where 16 bands were used.
With 16 bands there are more possibilities of transitions at higher energies, and so more peaks show up.
It is important to note the consistency and resilience of the results for the linear optical conductivity:
no changes for different energy cutoffs.

\begin{figure}[h]
 \centering
 \includegraphics[scale=0.25,keepaspectratio=true]{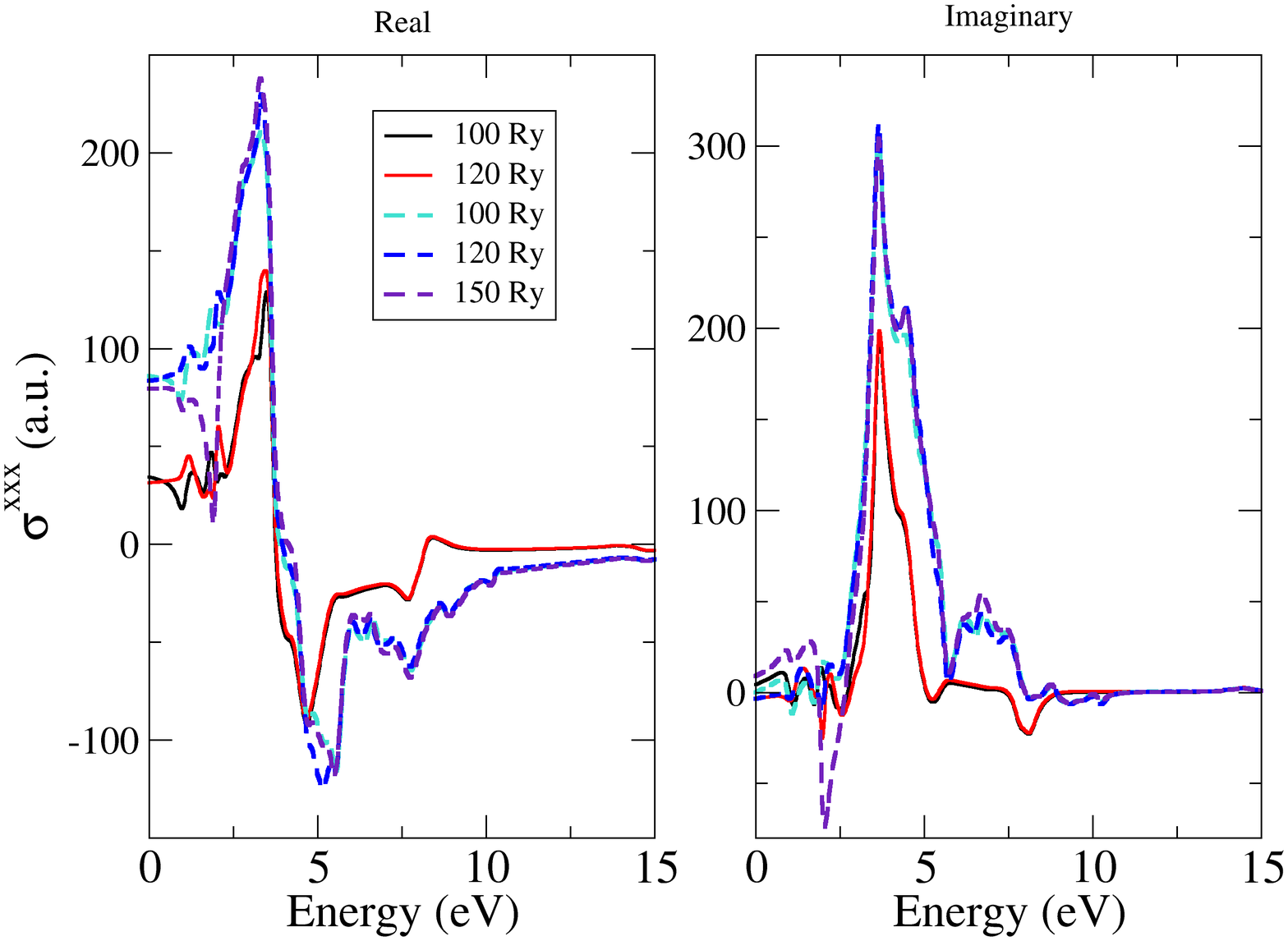}
 \caption{Second Harmonic Generation optical conductivity for InSe.
 Lines are for $47\times41$ $\pmb{k}$-points and dots are for $59\times51$ $\pmb{k}$-points.}
 \label{fig:sgh_inse}
\end{figure}

Figure \ref{fig:sgh_inse} shows the Second Harmonic Generation optical conductivity for InSe
for the five cases shown in Table \ref{tab:InSe}.

Again, there is a large difference between having more or less bands,
although the position of the main peaks is fairly accurate.
This time there are differences at low energies for different energy cutoffs.

\section{Computational Profile}
\label{sec:computational_profile}

The computational performance and limitations of {\sc berry} were assessed
using the single layer hexagonal boron nitride (hBN) material as a representative case study.
A variation of the analysis parameters,
such as the number of points in the reciprocal space meshes and the number of bands,
allows to manipulate the problem size, i.e. the amount of computation and/or data processed,
to evaluate {\sc berry}'s subroutines on different scenarios.

{\sc berry} was evaluated in a dual-socket server
with two 16-core Intel Xeon E5-2683v4 Broadwell devices @2.1 GHz
with 2-way simultaneous multithreading and 256 GiB RAM, connected to three types of storage:
a high-performance Network-Attached Storage (NAS) with HDDs and a SSD-based caching system over 10 Gbit ethernet,
common in cluster environments;
a local Crucial 1050 MX SSD; and a local Samsung 970 Evo Plus NVMe SSD.
Table \ref{tab:bandwidths} summarises the relevant performance metrics of the storage devices tested.
Bandwidth is more relevant for I/O interaction with larger amounts of data,
while latency affects the I/O of random accesses to small amounts of data.
The software was tested using CentOS 7.9.2 operating system, with Python 3.10 and Quantum Espresso 7.1.
A k-best measurement heuristic was used to ensure that the results can be replicated,
with k = 3 with a 5\% tolerance and a minimum/maximum of 10/20 measurements.

\begin{table}
\centering
\begin{tblr}{}
\hline
     & \SetCell[c=2]{c}{Read} & & \SetCell[c=2]{c}{Write} \\

     & Bandwidth (MiB/s) & Latency ($\mu$s) & Bandwidth (MiB/s) & Latency ($\mu$s) \\
     \hline
NAS  & 347,2      & 3771    & 158,2      & 4182      \\
SSD  & 415,3      & 689     & 470,4      & 892     \\
NVMe & 3559,2     & 295     & 3317,8     & 348    \\
\hline
\end{tblr}
\caption{Bandwidth of the different storage devices on the test server.}
\label{tab:bandwidths}
\end{table}

\subsection{Performance Assessment}
\label{subsec:performance}

\begin{figure}[!htb]
     \centering
     \begin{subfigure}[b]{0.49\textwidth}
         \centering
         \includegraphics[width=\textwidth]{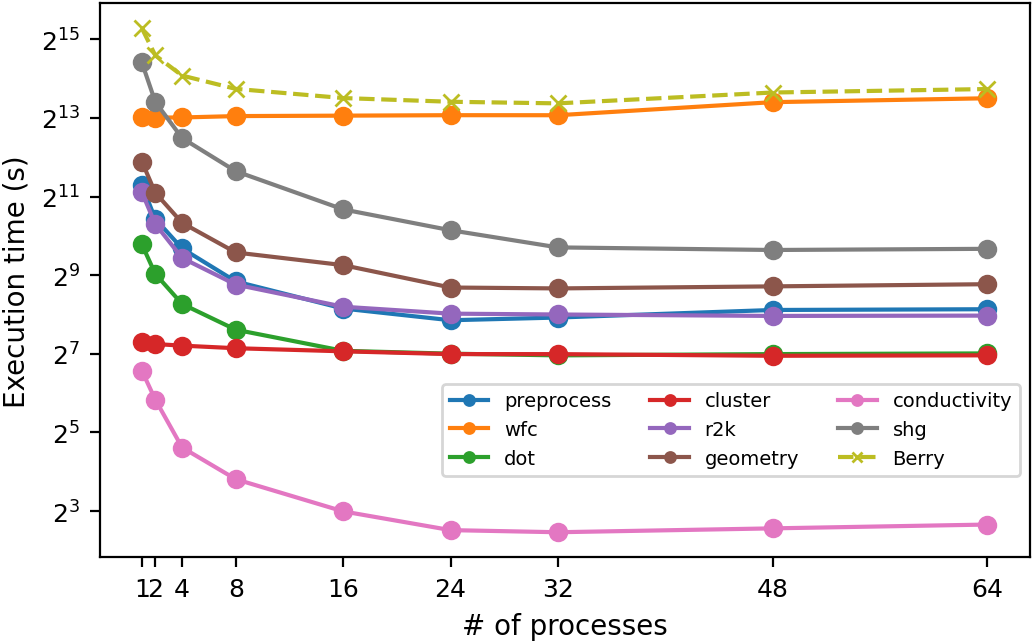}
     \end{subfigure}
     \hfill
     \begin{subfigure}[b]{0.49\textwidth}
         \centering
         \includegraphics[width=\textwidth]{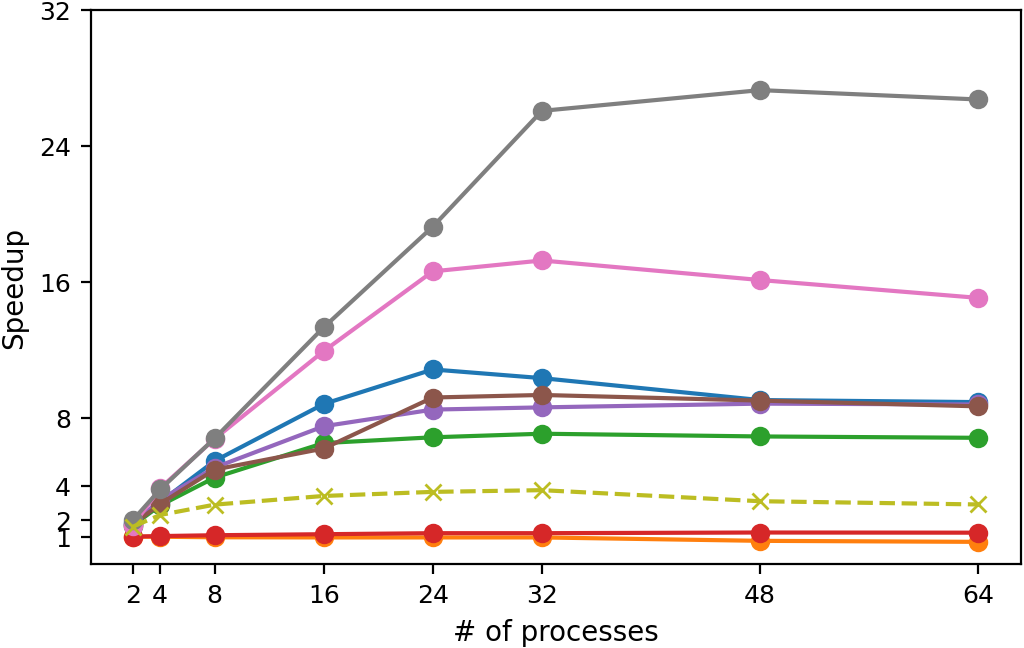}
     \end{subfigure}
        \caption{Execution time (left) and speedup (right) of {\sc berry}'s packages using up to 64 processes.
        Speedup of $n$ processes is measure relative to the sequential run.}
        \label{fig:exec_times}
\end{figure}

Figure \ref{fig:exec_times} presents the individual performance of {\sc berry}'s packages
on the test server using the NAS storage over Ethernet,
varying the number of parallel processes used to assess its scalability.
The hBN material is used with $24\times24$ $\pmb{k}$-points, 20 bands, and an energy cutoff of 100 Ry.
The \texttt{wfc} and \texttt{shg} aggregate processing accounts between 76\% and 89\% of the overall execution time,
for 1 and 32 processes respectively.
This relative increase is due to \texttt{wfc} performance not being affected
by the increased amount of parallel processes used,
since this package is limited by the storage I/O performance
(it performs a small amount of computation per byte read/written).
Similar behaviour is shown by \texttt{cluster} and \texttt{basis},
but with a less significant impact on the overall performance due to their faster execution times,
accounting at most for 1.2\% of {\sc berry}'s time.
An overview of the relative execution time of each of {\sc berry}'s packages using 1 and 32 processes is presented in Table \ref{tab:exec_percentage}.

\begin{table}[!htb]
\begin{tabular}{lccccc}
\hline
                      & \textbf{preprocess} & \textbf{wfc}      & \textbf{dot}          & \textbf{cluster} & \textbf{basis} \\ \hline
\textbf{sequential}   & 6,31\%              & 21,07\%           & 2,21\%                & 0,39\%           & 0,01\%         \\ \hline
\textbf{32 processes} & 2,29\%              & 81,13\%           & 1,18\%                & 1,20\%           & 0,02\%         \\ \hline
\textbf{}             & \textbf{r2k}        & \textbf{geometry} & \textbf{conductivity} & \textbf{shg}     & \textbf{}      \\ \hline
\textbf{sequential}   & 5,54\%              & 9,53\%            & 0,24\%                & 54,70\%          &                \\ \hline
\textbf{32 processes} & 2,42\%              & 3,83\%            & 0,05\%                & 7,89\%           &                \\ \hline
\end{tabular}
\caption{Relative execution time of each of {\sc berry}'s packages.}
\label{tab:exec_percentage}
\end{table}

\texttt{shg} and \texttt{conductivy} are the best scaling packages,
improving the performance over the sequential code by $28\times$ and $17\times$ using 32 processes.
This would be expected as they are the most computational intensive packages of {\sc berry}.
\texttt{r2k}, \texttt{preprocess}, \texttt{geometry}, and \texttt{dot} exhibit less significant scalability,
with peak speedups around $8.5\times$, $10.8\times$, $9.2\times$, and $6.9\times$, respectively, for 24 and 32 processes.
However, none of these packages effectively use the extra computing resources,
with parallelization efficiencies below 50\%, with the exception of \texttt{shg}.
\texttt{cluster}, and \texttt{wfc} do not scale with the number of processes
as they are significantly bound either memory (latency or bandwidth) or I/O performance,
thus not benefiting by the extra processes used.
None of the packages benefit from using more processes than the available physical cores in the server.

\begin{figure}[!htb]
     \centering
     \begin{subfigure}[b]{0.49\textwidth}
         \centering
         \includegraphics[width=\textwidth]{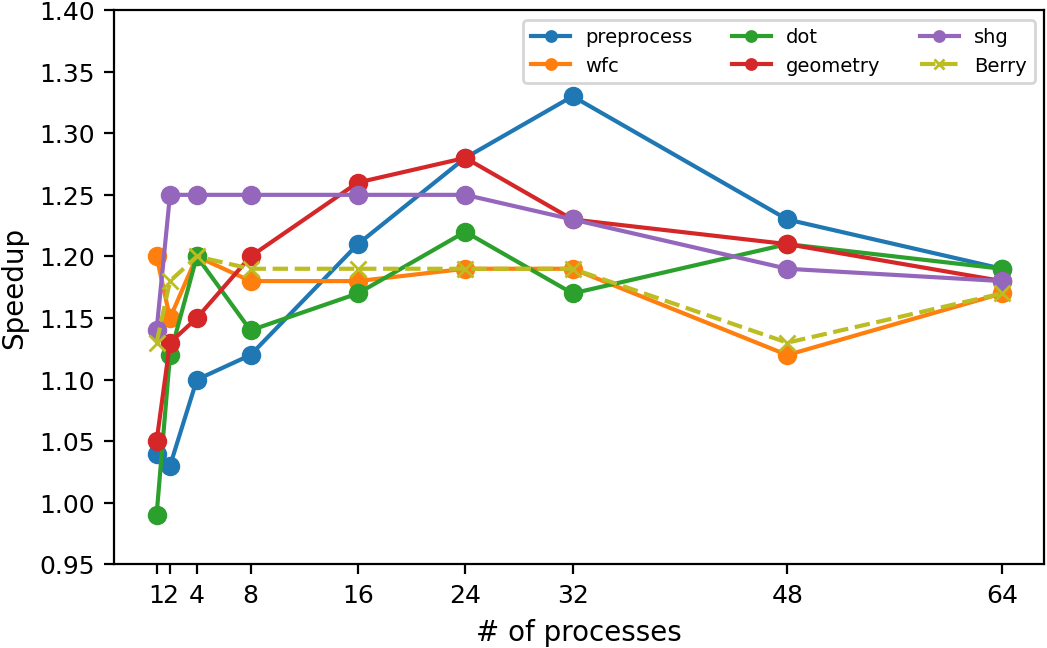}
     \end{subfigure}
     \hfill
     \begin{subfigure}[b]{0.49\textwidth}
         \centering
         \includegraphics[width=\textwidth]{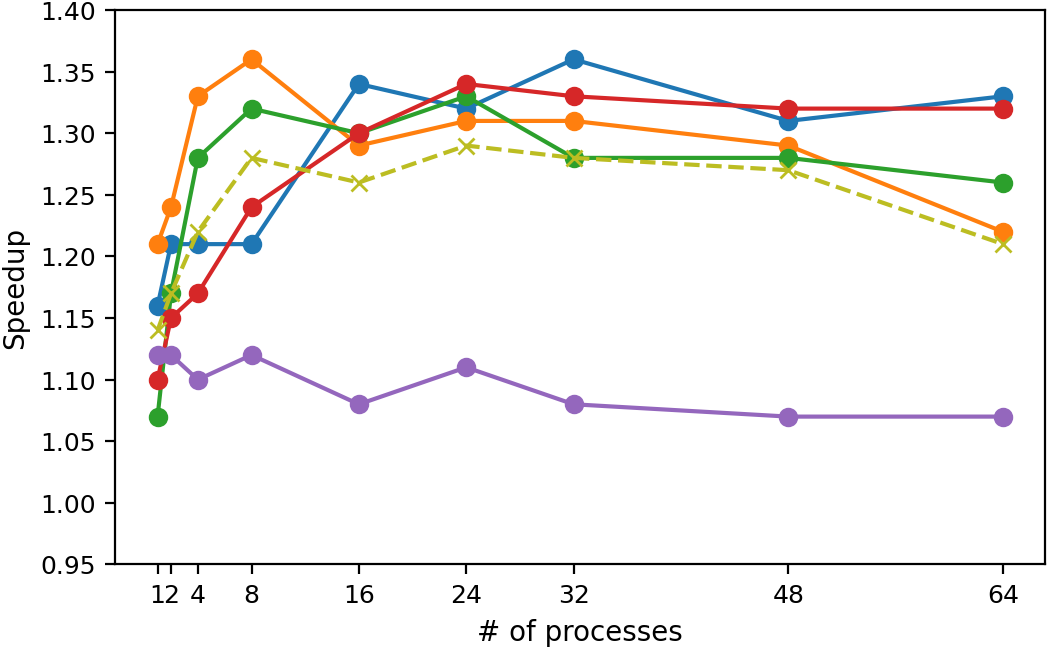}
     \end{subfigure}
        \caption{Speedup of the SSD \textit{vs} NAS (left) and the NVMe \textit{vs} SSD (right) storage devices for each of {\sc berry}'s packages using up to 64 processes.}
        \label{fig:storage_comparison}
\end{figure}

Figure \ref{fig:storage_comparison} shows the impact of using a faster SSD or NVMe storage
on the performance of {\sc berry}'s packages.
\texttt{r2k}, \texttt{cluster}, and \texttt{conductivity} are omitted as they did not significantly benefit
from the improved storage, as they are compute-bound and have small execution times.
The remaining packages
benefit from using an SSD over NAS mostly due to the lack of performance of the traditional HDDs in the NAS,
with improvements generally between 10\% and 30\%.
\texttt{shg} benefit less from using a NVMe over a SSD
as the bump in performance provided by the SSD ensures they are mostly compute-bound.
\texttt{preprocess} and \texttt{wfc} greatly benefit from both SSD and NVMe devices,
due to their bandwidth intensive operations (large amounts of sequential data being read and written),
especially when using a higher amount of processes.
\texttt{dot} and \texttt{geometry} benefit the most from the NVMe due to its latency,
as these packages have more random memory access patterns with less data being required per access.

The overall efficiency of {\sc berry}'s parallelization is relatively low since it is mostly limited by \texttt{wfc}, whose performance does not scale with the number of processes, making it memory-bound when considering all packages.
Significant performance improvements may be possible by tackling the I/O bottlenecks limiting this package, especially when using more than 4 processes where {\sc berry}'s parallelization efficiency drops below 50\%.
The use of faster storage also provided improvements for compute-bound packages due to the contention when writing the output data using large number of processes, which may be especially problematic when using HDD-based NAS.

\subsection{Performance Scalability with Problem Size}
\label{subsec:scalability}

This subsection focuses on the analysis of the impact that different input data size and configuration has on the execution time and size of the output data of {\sc berry},
using 32 processes in the test server and SSD storage.

\begin{figure}[!htb]
     \centering
     \begin{subfigure}[b]{0.49\textwidth}
         \centering
         \includegraphics[width=\textwidth]{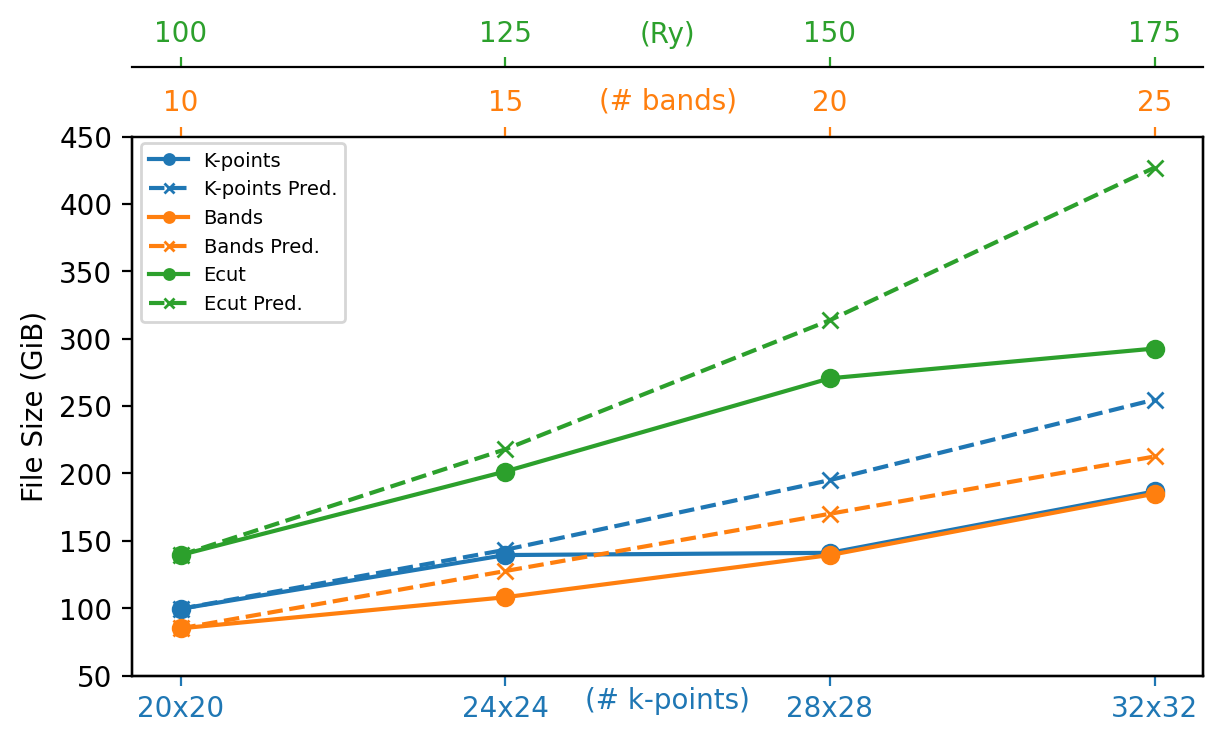}
     \end{subfigure}
     \hfill
     \begin{subfigure}[b]{0.49\textwidth}
         \centering
         \includegraphics[width=\textwidth]{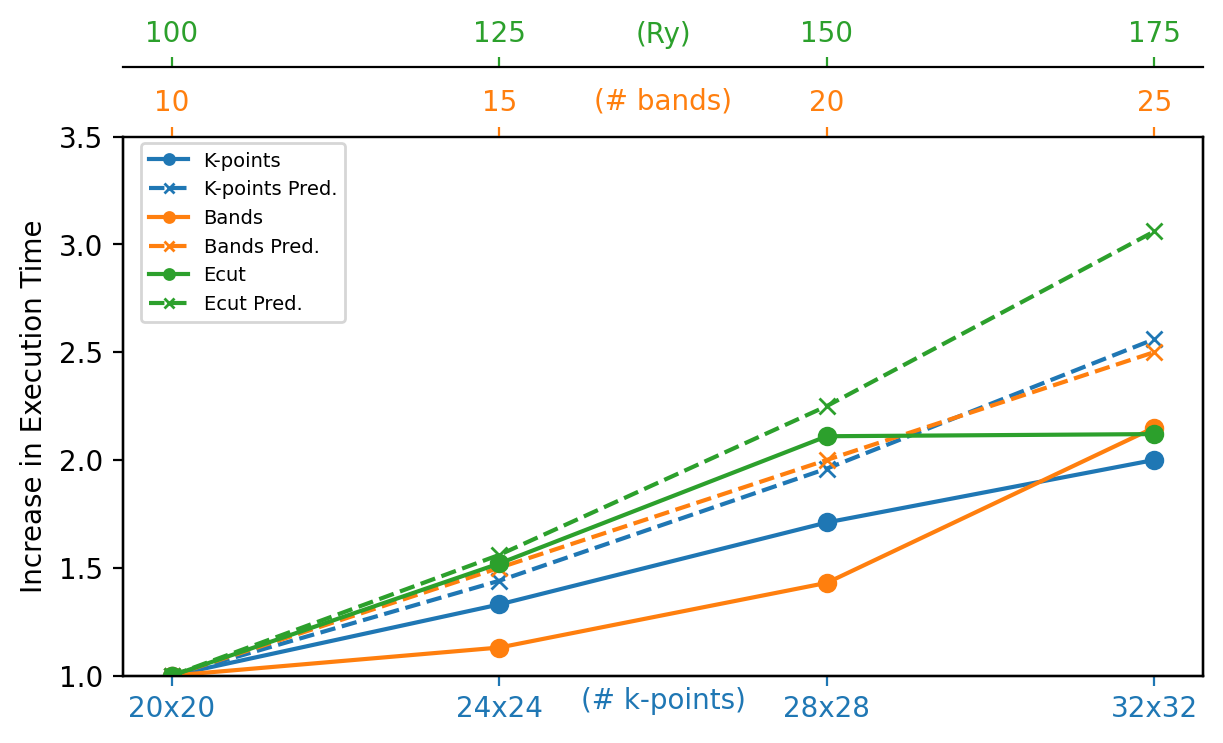}
     \end{subfigure}
        \caption{{\sc berry}'s predicted (\textit{Pred}) and measured execution time (left) and output data size (right) for different problem sizes.}
        \label{fig:scalability}
\end{figure}

Figure \ref{fig:scalability} presents predictions and measurements of the increase in {\sc berry}'s execution time, compared to a reference configuration,
and storage requirements for three different problem sizes using the hBN material:
(i) varying the number of $\pmb{k}$-points, which has a linear impact on computation;
(ii) varying the number of bands to be considered, which also has a linear impact on computation;
and (iii) the energy cutoff for the electrons, which should quadratically increase the computing and storage of its data.

Duplicating the number of bands only increases the execution time by 40\%,
which is significantly below the predicted 2x increase.
This behaviour is expected since the amount of bands affects especially the \texttt{cluster} package,
which is compute-bound but only accounts for roughly 1\% of {\sc berry}'s execution time,
and other memory-bound packages to a lower degree, thus mitigating the extra computing required.
However, the required storage increases almost as expected.
The execution time for the increase in $\pmb{k}$-points also falls short of the prediction
even though the amount of computation required is almost doubled,
which is also due to the memory-bound nature of {\sc berry}.
The output size increase is almost inline with expectations
(7\% short of the prediction for $32\times32$ $\pmb{k}$-points).
A 50\% increase in the energy cutoff, from 100 to 150 Ry, results in a quadratic increase in computing time and storage, falling short of the predicted cube increase in both metrics.
The shorter increase in {\sc berry}'s execution time is again due to it being memory-bound that mitigates the increase in computations.
In terms of storage, there is a cubic increase in the data output by \texttt{wfc}, whose effect is diluted by other non-scaling data outputs.
Between 150 and 175 Ry, the execution time barelly changes; this is because increasing the energy cutoff increases the numerical
accuracy of the DFT calculations, and this results in a faster convergence in some programs, specially the \texttt{cluster} package.
This compensates the increase in time due to the larger ammount of calculations in other packages.

These analyses of {\sc berry}'s performance can be valuable from the perspective of an user,
where the improvements in execution time do not necessarily mean that the time-to-solution will decrease,
but rather allow for more complex and/or larger problems to be tackled \cite{gustafson,gustvsamd}.
Assuming the extrapolations of the data in Figure \ref{fig:scalability}
and considering the best performing configuration presented in subsection \ref{subsec:performance},
{\sc berry} is capable of processing either $16\times16$ $\pmb{k}$-points, 20 bands more, or an energy cutoff twice higher when using 32 processes within the time frame of a sequential run.

\section{Future perspectives}
\label{sec:future}

We recognise that {\sc Berry} can be improved in both functionality and performance.

The inefficient parallelization of specific packages can be improved,
especially targeting I/O- and memory-bound tasks,
as it is not currently feasible to analyse 3 dimensional materials in a reasonable time frame.
Such improvements should be detached from {\sc Quantum ESPRESSO} to ensure compatibility
with previous and future versions of this software.
There will also be improvements in the determination of bands.

Another important feature is to be able to use noncolinear DFT calculations.
This will allow a large expansion on the number of materials that can be dealt with,
in a rigurous manner.

Other properties will be implemented, like excitons and Hall conductivity,
and higher order non-linear optical properties.

\section{Conclusions}
\label{sec:conclusions}

We developed a software that successfuly generates sets of eigenstates that are differentiable
in reciprocal space, departing from purely numerical density functional calculations.

Several hurdles had to be overcome to achieve this, namely establishing
 a procedure to solve the problem of random phases in numerical calculations of electronic structure
due to the independence of basis at different $\pmb{k}$ points,
and thus making all wavefunctions coherent;
finding degenerate points and correcting the basis to achieve continuity;
and sorting out the eigenstates in analytic sets. This last step involved the use of Graph Theory
and Unsupervised Machine Learning in an iterative process.

The solution allowed us to calculate the gradients of the Bloch factor in reciprocal space,
and the Berry connections and curvatures from first principles.
We then used this knowledge for the calculation of optical response
in linear order and the second harmonic generation in hBN and InSe.

This software opens huge possibilities in the study of responses in 2D crystals
as well as other properties that may be calculated from the eigenstates
along bands in reciprocal space.

\section*{Acknowledgments}

We acknowledge support by the Portuguese Foundation for Science
and Technology (FCT) in the framework of the Strategic Fundings UIDB/04650/2020, UIDB/50007/2020 and
Associate Laboratory LA/P/0016/2020 and COMPETE 2020, PORTUGAL 2020, FEDER and FCT through project
 PTDC/FIS-MAC/2045/2021 and from the European Commission through the project Graphene
Driven Revolutions in ICT and Beyond (Ref. No. 881603, CORE 3).





\bibliographystyle{elsarticle-num}

\bibliography{paper_berry}







\end{document}